\documentclass[manuscript]{aa}

\usepackage{amsmath}
\usepackage{graphicx}
\usepackage{txfonts}

\begin{document}

\title{Evidence for a Toroidal Magnetic-Field Component in 5C4.114 on
Kiloparsec Scales}

\author{
Denise C. Gabuzda,\inst{1}
Sebastian Knuettel,\inst{1}
\and
Annalisa Bonafede\inst{2}
 }

\institute{
Department of Physics, University College Cork, Cork, Ireland. \\
                 Email: d.gabuzda@ucc.ie, s.knuettel@mars.ucc.ie\\
\and Hamburger Sternwarte, Gojenbergsweg 112, 21029 Hamburg, Germany\\
                 E-mail: annalisa.bonafede@hs.uni-hamburg.de
}

\def\gsim{\mathrel{\raise.5ex\hbox{$>$}\mkern-14mu
             \lower0.6ex\hbox{$\sim$}}}

\def\lsim{\mathrel{\raise.3ex\hbox{$<$}\mkern-14mu
             \lower0.6ex\hbox{$\sim$}}}

\date {Received ; accepted }

\abstract
{
A monotonic, statistically significant gradient in the observed Faraday
Rotation Measure (RM) across the jet of an Active Galactic Nucleus (AGN) 
reflects a corresponding gradient in the electron density and/or line-of-sight
magnetic (B) field in the region of Faraday rotation. For this reason,
such gradients may indicate the presence of a toroidal B field
component, possibly associated with a helical jet B field.
Although transverse RM gradients have been reported across a number of 
parsec-scale AGN jets, the same is not true on kiloparsec scales, 
suggesting that other (e.g. random) magnetic-field components
usually dominate on these larger scales.
}
{We wished to identify clear candidates for monotonic, transverse
RM gradients across AGN jet and lobe structures on scales larger than 
those probed thus far, and estimate their statistical significances. 
}
{We identified an extended, monotonic transverse Faraday rotation gradient
across the Northern lobe of a previously published Very Large Array 
(kiloparsec-scale) RM image of 5C4.114. We reanalyzed these
VLA data in order to determine the significance of this RM gradient.
}
{
The RM gradient across the Northern kiloparsec-scale lobe structure
of 5C4.114 has a statistical significance of about $4\sigma$. There is also 
a somewhat less prominent monotonic transverse Faraday rotation 
gradient across the Southern jet/lobe (narrower range of distances from the 
core, significance $\simeq 3\sigma$).
Other parts of the Faraday Rotation distribution observed
across the source are patchy and show no obvious order.
}
{
This suggests that
we are observing a random RM component associated with the
foreground material in the cluster in which the radio source is located
and through which it is viewed, superposed on
a more ordered RM component that arises in the immediate
vicinity of the AGN jets.  We interpret the transverse RM 
gradient as reflecting the systematic variations of the line-of-sight 
component of a helical or toroidal B field associated with the jets
of 5C4.114. These results suggest that the helical field that arises due
to the joint action of the rotation of the central black hole and its
accretion disc and the jet outflow can survive to distances of thousands
of parsec from the central engine.
}


\authorrunning{Gabuzda et al. 2015}
\titlerunning{Toroidal Magnetic-Field Component in 5C4.114}

\maketitle

\section{Introduction}

The radio emission of radio galaxies and Active Galactic Nuclei (AGNs)
is synchrotron emission, which can be linearly polarized up to about
75\% in optically thin regions, where the polarization angle
$\chi$ is orthogonal to the projection of the magnetic field
{\bf B} onto the plane of the sky (Pacholczyk 1970). Linear 
polarization measurements thus
provide direct information about both the degree of order and the
direction of the {\bf B} field giving rise to the observed synchrotron
radiation.

Multi-frequency interferometric polarization observations also provide
high-resolution information about the distribution of the Faraday 
rotation of the observed polarization angles arising between the 
source and observer.  When the Faraday rotation
occurs outside the emitting region in regions of non-relativistic
plasma, the amount of rotation is given by
\begin{eqnarray}
           \chi_{obs} - \chi_o =
\frac{e^3\lambda^{2}}{8\pi^2\epsilon_om^2c^3}\int n_{e}
{\mathbf B}\cdot d{\mathbf l} \equiv RM\lambda^{2}
\end{eqnarray}
where $\chi_{obs}$ and $\chi_o$ are the observed and intrinsic
polarization angles, respectively, $-e$ and $m$ are the charge and
mass of the particles giving rise to the Faraday rotation, usually
taken to be electrons, $c$ is the speed of light, $n_{e}$ is the
density of the Faraday-rotating electrons, $\mathbf{B} \cdot
d\mathbf{l}$ is an element of the line-of-sight magnetic field,
$\lambda$ is the observing wavelength, and RM (the coefficient of
$\lambda^2$) is the Rotation Measure (e.g., Burn 1966).  Simultaneous
multifrequency observations thus allow the determination of both
the RM, which carries information about the electron density and
{\bf B} field in the region of Faraday rotation, and $\chi_o$,
which carries information about the intrinsic {\bf B}-field geometry
associated with the synchrotron source.

Systematic gradients in the Faraday rotation have been reported
across the parsec-scale jets of a number of AGN, interpreted as
reflecting the systematic change in the line-of-sight component of
a toroidal or helical jet {\bf B} field across the jet (e.g., 
Hovatta et al. 2012, Mahmud et al. 2013, and Gabuzda et al.
2014b, 2015).  Such fields would come about in a natural way as a result
of the ``winding up'' of an initial ``seed'' field by the differential
rotation of the central accreting objects (e.g. Nakamura, Uchida \&
Hirose 2001; Lovelace et al. 2002).

It is an interesting question whether these ordered fields can survive to
larger scales as the jets propagate outward. The presence of a clear
transverse RM gradient on scales of more than a hundred parsec from the jet
base was reported for 3C~380 (Gabuzda et al. 2014a). However, attempts to
search
for possible kiloparsec-scale transverse Faraday-Rotation gradients across
the jet structures of extragalactic radio sources by eye have yielded only
a small number of candidates (Gabuzda et al. 2012). This is consistent with
a picture in which random distributions of the {\bf B} field and electron
density in the general vicinity of the radio source (e.g., in the cluster or
inter-cluster medium in which the source is located) generally dominate on
these large scales. Only one claim of a transverse Faraday-rotation gradient
that could be associated with a helical {\bf B} field present on kiloparsec
scales has been made (Kronberg et al. 2011).

We present here a new analysis of observations of the radio galaxy
5C~4.114, whose position is approximately coincident with the location of
the Coma cluster of galaxies, originally obtained and analyzed by Bonafede 
et al. (2010), which
demonstrates the presence of monotonic, statistically significant
Faraday Rotation gradients across both the Northern lobe 
and Southern jet/lobe. The
Northern gradient is especially striking. If associated
with the azimuthal components of helical {\bf B}  fields, the orientation
of the gradients is consistent with the initial  {\bf B}  field
that is wound up having a dipolar-like configuration.

\section{Observations and Reduction}

The data used for our analysis are precisely the 1.365, 1.516, 4.535 
and 4.935~GHz Very Large Array data considered by Bonafede et al. (2010), 
and the observations and the data calibration and
reduction methods used are described in that paper. 

We could not use the RM map published by Bonafede et al. (2010) directly,
because the associated error map did not take into account the finding of
Hovatta et al (2012) that the uncertainties in the Stokes $Q$ and 
$U$ fluxes in individual pixels on-source are somewhat higher than 
the off-source rms fluctuations, potentially increasing the resulting
RM uncertainties. 

To address this, we imported the final, fully self-calibrated
visibility data of Bonafede et al. (2010) into the {\sc AIPS} package, then
used these data to make naturally weighted $I$, $Q$ and $U$ maps
at all four frequencies, with matching image sizes, cell sizes and beam
parameters specified by hand in the {\sc AIPS} task {\sc IMAGR}.
These images were all convolved with a circular Gaussian beam having a
full-width at half-maximum of 1.3$^{\prime\prime}$. We then obtained
maps of the polarization angle, $\chi = \frac{1}{2}\arctan(U/Q)$, and
used these to construct corresponding maps of the Faraday Rotation Measure
(RM) using the {\sc AIPS} task {\sc RM}.  The uncertainties in the polarization
angles used to obtain the RM fits were calculated from the uncertainties in $Q$
and $U$, which were estimated using the approach of Hovatta et al. (2012).
The output pixels in the RM maps were blanked when the RM uncertainty
resulting from the $\chi$ vs. $\lambda^2$ fits exceeded 8~rad/m$^2$.

\section{Results}

Fig. 1 presents our 4.535-GHz intensity image of 5C4.114, with the RM image
superposed in colour; these essentially reproduce the images in Fig.~12 
of Bonafede et al. (2010). 

\begin{figure*}
\begin{center}
\hspace*{-1.0cm}
\includegraphics[width=.65\textwidth,angle=-90]{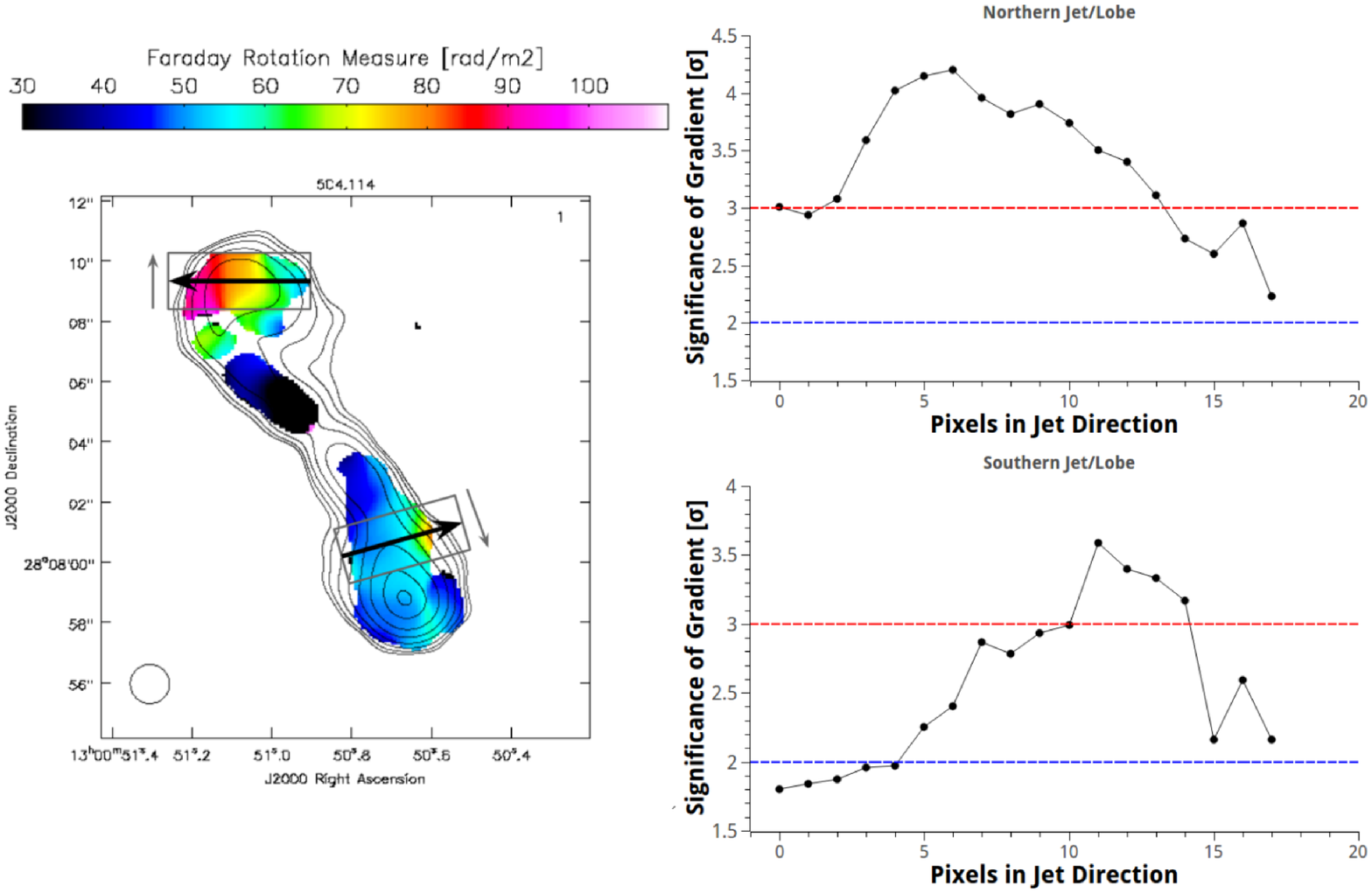}
\end{center}
\hspace*{0.3cm}
\vspace*{-1.0cm}
\caption{Left: 4.9-GHz intensity map of the 5C4.114 with the RM 
distribution
superposed. The intensity peak is 3.6~mJy/beam, the lowest intensity
contour is $5.0\times 10^{-2}$~mJy/beam, and the contours increase in
increments of a factor of two.  The two black arrows highlight transverse RM
gradients visible across the Northern lobe and the Southern jet/lobe.  The 
grey boxes show the regions for which the significances of series of 
parallel, monotonic transverse RM gradients are plotted in the right-hand 
panels; the grey arrows pointing outward along the jet show the direction 
of increasing pixel number in the right panels, and pixel~0 corresponds 
to the inner edge of each of the grey boxes.  The pixel size is 0.1~arcsec.  
The $2\sigma$ level is shown by the dashed blue horizontal lines, and the 
$3\sigma$ level by the dashed red horizontal lines.}
\end{figure*}

A monotonic gradient in the RM across the Northern radio lobe 
is clearly visible by eye, highlighted by the upper black arrow in 
Fig.~1. A less prominent, oppositely directed
RM gradient is visible across
the Southern jet/lobe, highlighted by the lower black arrow in Fig.~1.
Both regions are fairly well resolved, and span two to three beamwidths in 
the transverse direction. The ordered RM gradient crossing the Northern
lobe is quite unusual for the kiloparsec-scale Faraday rotation distributions
of radio galaxies and quasars, which tend to be more patchy.


The redshift
of 5C4.114 is not known, but no optical identification with either a Coma
cluster galaxy nor with a background galaxy has been found, indicating that
the source's redshift is greater than 0.023 (Bonafede et al. 2010). This
indicates that the projected distance from the AGN core to the location
of the Northern transverse RM gradient is at least 2~kpc (assuming a
Hubble constant $H_o = 71$~km\,s$^{-1}$Mpc$^{-1}$, $\Omega_{\Lambda} =
0.73$, and $\Omega_m = 0.27$), with the projected
distance to the Southern gradient being somewhat smaller.

\subsection{Significance of the Transverse RM Gradients}

Monotonic transverse RM gradients are observed throughout the
regions enclosed by the grey boxes in Fig.~1; each of the points in 
the right-hand panels of Fig.~1 corresponds to a monotonic transverse
RM gradient at some distance from the core.
The uncertainties of the RM values were determined using 
$\chi$ uncertainties estimated in individual
pixels using the approach of Hovatta et al. (2012), without including the
effect of uncertainty in the EVPA calibration, since this cannot introduce
spurious RM gradients (Mahmud et al. 2009, Hovatta et al. 2012).  The
uncertainty of the difference between the RM values at the two ends of
a slice was estimated by adding the uncertainties for the two RM values
in quadrature.  Comparisons of the RM values at the two ends of the 
RM slices considered in Fig.~1 indicate that the Northern and Southern 
transverse gradients have significances reaching $4.2\sigma$ and 
$3.6\sigma$, respectively.  

The significance of the transverse RM gradients across the Northern lobe 
(upper right plot in {\bf Fig.~1}) is $\simeq 3-4.2\sigma$ throughout 
nearly the entire region considered, and everywhere exceeds $2\sigma$. 
This is thus a quite extended and coherent RM structure. The significance
of the less prominent transverse RM gradients across the Southern 
jet/lobe (lower right plot in Fig.~1) is greater than $2\sigma$ 
throughout most of the region enclosed by the Southern grey box in Fig.~1, 
and reaches $\simeq 3-3.6\sigma$ in the vicinity of the transverse 
black arrow in Fig.~1. 

\section{Discussion}

\subsection{Ability to Detect the Transverse RM Gradients}

It is important to note here that we are interested in establishing whether
the transverse RM gradients visible by eye represent a statistically 
significant, systematic, 
monotonic change in the observed Faraday RM across the lobe/jet structure. 
We are not interested in measuring the intrinsic RM values, only in 
investigating the reality of the observed gradients.

The question of the resolution necessary to reliably detect the presence
of a transverse RM gradient has been discussed in the literature in the 
context of VLBI-scale RM measurements. Taylor \& Zavala (2010) had proposed 
that the reliable detection of a transverse RM gradient required that the
observed RM gradient span at least three ``resolution 
elements'' (usually taken to mean three beamwidths) across the jet.
This was tested
by Mahmud et al. (2013) using Monte Carlo simulations based on model 
core--jet-like sources with transverse RM gradients present across their 
structures, which had  intrinsic jet widths of 1/2, 1/3, 1/5, 1/10 and 
1/20 of the beam full-width at half-maximum (FWHM) in the direction across 
the jet.  The 
resulting simulations show that {\em the transverse RM gradients introduced 
into the model visibility data remained visible in RM maps constructed from 
realistic ``noisy'' data using standard techniques, even when the intrinsic 
width of the jet structure was much smaller than the beam width.} 


These Monte Carlo simulations (and also those of Hovatta et al.
(2012)) thus directly demonstrate that
the three-beamwidth criterion of Taylor \& Zavala (2010) is overly restrictive,
and that it is not necessary or meaningful to place a limit on the width
spanned by an RM gradient in order for it to be reliable: the key criteria
are that the gradient be monotonic and that the difference between the 
RM values at either end be at least $3\sigma$.
The counterintuitive 
result that a transverse RM gradient spanning only one beamwidth can 
potentially be significant essentially comes about because polarization is
a vector quantity, while the intensity is a scalar. 
Thinking of the polarization
as being composed of Stokes $Q$ and $U$, this enhanced sensitivity to
closely spaced structures comes about because both $Q$ and $U$ can be
positive or negative. Again, we are not speaking here of being able to 
accurately deconvolve the observed RM profiles to determine the intrinsic 
transverse RM structure --- only of the ability to detect the presence 
of a systematic transverse RM gradient.

This means that we can consider the presence of the RM gradients detected 
across the Northern lobe and Southern jet/lobe of 5C4.114, which are
monotonic, encompass a range of RM values of at least $3\sigma$, and span 
$\simeq 2-3$~beamwidths across the jet and $\simeq 0.5-1.5$~beamwidths
along the jet, to be reliably detected, although it is not possible to 
derive the intrinsic (infinte-resolution) values of the gradients. The
three-dimensional structures of the emission regions where the RM gradients
are observed cannot be determined with certainty, although the general
symmetry of the radio structure observed suggests that both the Northern 
and Southern jet/lobe structures are not very far from the plane of the sky.

\subsection{Random or Ordered RM Distributions on kpc Scales?}

It is generally believed that the RM distributions across 
extragalactic radio sources on kiloparsec scales should be dominated
by fairly random distributions of the electron density and {\bf B}  field in
plasma that is not closely related to the radio source itself, for example,
plasma that is associated with the cluster or inter-cluster medium. This view
is supported by the fact that most of the observed RM distributions appear
irregular and ``patchy'' (e.g., the RM maps presented by Bonafede et
al. (2010) and Govoni et al. (2010)). This was the framework in which the
original study of Bonafede et al. (2010), aimed at constraining the Coma
cluster magnetic-field strength, was carried out.

Based on their results for the seven FRI radio sources they considered,
Bonafede et al. (2010) concluded that the observed RM distributions 
generally
did not originate in the immediate vicinity of the sources, arising instead
in intervening material of the Coma cluster.  This conclusion was based
on statistical analyses of the RMs and their uncertainties for different
lines of sight through the cluster, and did not take into consideration
possible patterns in the RM distributions for the individual objects studied.

\begin{figure*}
\begin{center}
\includegraphics[width=.15\textwidth,angle=90]{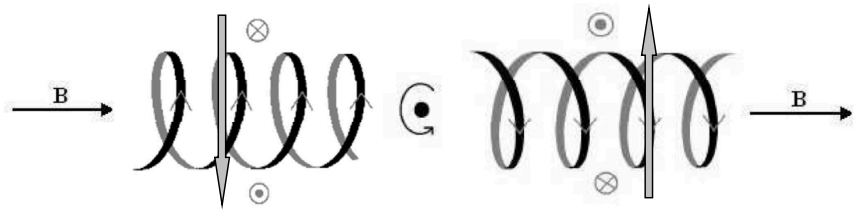}
\end{center}
\caption{Schematic of the effect of winding up a dipole-like poloidal
{\bf B}  field (shown by the vertical arrows) by rotation of the central
black hole and accretion disk.  The circled dots and circled X's show the
resulting azimuthal {\bf B}  field oriented towards and away from the
observer on either side of the jets, and the filled grey arrows show the
direction of the associated RM gradients.}
\end{figure*}

However, the conclusions of Bonafede et al. (2010) for their sample 
as a whole and our own conclusions specifically for 5C4.114 need
not be in contradiction. We suggest that, indeed, the observed RM distributions
of extragalactic radio sources on kiloparsec scales are usually (though not
always) determined
by intervening magnetised plasma that is not directly related to the jets;
this plasma is not uniform and possibly turbulent, giving rise to an
irregular, patchy RM distribution that bears no obvious relationship to the
source structure. However, the resulting irregular RM distribution can be
superposed on a more ordered RM distribution brought about by magnetised
plasma in the immediate vicinity of the jet structure. Although it is usually
the irregular RM component that is dominant, the ordered RM component may
occasionally be visible in some individual objects. We suggest that this is
the case for 5C4.114.

\subsection{Tentative Evidence for a Dipole-like Initial Field Structure}

The transverse RM gradient detected across the 
Southern jet/lobe structure of 5C4.114 is somewhat less prominent
than the Northern RM gradient, which is very 
firmly detected. In this section,
we explore the implications of the overall RM-gradient structure if both the 
(firm) Northern and (somewhat tentative) Southern RM gradients reflect
the azimuthal (toroidal)
component of a helical jet {\bf B}  field, brought about by the combination
of the rotation of the central black hole and its accretion disk and the
jet outflow. In this case, the directions of the transverse RM gradients on
the sky imply particular directions for the associated azimuthal {\bf B}-field
components; this, in turn, enables us to infer whether the poloidal components
of the ``seed fields'' that were wound up by the rotation of the central
black hole and accretion disk had the same or opposite senses in the two jets.

The direction of the azimuthal field component is determined by the direction
of the central rotation and the direction of the poloidal component of the
initial field that is ``wound up''.  As is illustrated schematically in
Fig.~3, the pattern shown by 5C4.114 is consistent with the poloidal
{\bf B}-field component being directed outward in one jet and inward in
the other, as would be expected for a dipolar-like initial {\bf B}-field
configuration.

\section{Conclusion}

We have reconstructed the Faraday RM image of 5C~4.114 initial published
by Bonafede et al. (2010) in order to
quantitatively analyze various gradients visible in the RM image.

Our analysis has demonstrated that the differences in the RM values 
encompassed by the monotonic RM gradients visible across the entire 
Northern lobe of the radio source and a more restricted
region in the Southern jet/lobe both exceed $3\sigma$, making 
them statistically significant. The detection of the RM gradient across
the Northern lobe is very firm, while the RM gradient across the Southern 
jet/lobe is slightly more tentative, due to the relatively narrow range 
of distances from the central AGN where the significance of the gradient 
exceeds $3\sigma$.

This represents firm evidence that the Northern, and possibly
also the Southern, kiloparsec-scale jet of 5C4.114 carries a helical or 
toroidal {\bf B}-field component. 
Such a component would naturally arise
due to the rotation of the central black hole and its accretion disc;
apparently, this helical {\bf B}-field component can sometimes
survive to distances of thousands of parsec from the central engine.
Regardless of whether the regions where the RM gradients are observed
represent outwardly propagating jets or lobes of material flowing
backward toward the center of activity, they could contain the imprint of
a helical {\bf B}-field component that was initially carried outward by the
jet outflow.
The relative orientations of the Northern and Southern gradients
are consistent with the pattern expected if the initial poloidal
jet {\bf B} field had a dipole-like structure.

Our new analysis together with the original analysis of Bonafede et al.
(2010) suggest a picture in
which the observed RM distributions of extragalactic radio sources on
kiloparsec scales are usually determined by intervening inhomogeneous,
possibly turbulent magnetised plasma that is not directly related to the
jets, with a more ordered RM distribution associated with magnetised plasma
in the immediate vicinity of the jet structure occasionally becoming
dominant in some regions of individual sources. This latter, ordered
RM component can bear the imprint of a helical magnetic field associated
with the jet structure.


\end{document}